# Practical Approach of Knowledge Management in Medical Science

M. Bohlouli[1], P. Uhr[1], F. Merges[1], S. Mohammad Hassani[1], and M. Fathi[1]

[1] Department of Computer Science and Electrical Engineering, University of Siegen, NRW, Germany

**Abstract** - *Knowledge organization, infrastructure, and knowledge based activities are all subjects which help in creation of business strategies for new enterprise. In this paper, first basics of knowledge based systems are studied. Practical issues and challenges of Knowledge Management (KM) implementations are then illustrated. Finally comparison of different knowledge based projects is presented along with abstracted information on their implementation, techniques and results. Most of these projects are in the field of medical science. Based on our study and evaluation on different KM projects, we conclude that KM is being used in every science, industry and business. But its importance in medical science and assisted living projects is highlighted nowadays with the most of research institutes. Most medical centers are interested in using knowledge based services like portals and learning techniques of knowledge for their future innovations and supports.*

**Keywords:** *Knowledge Management, Knowledge Based Systems, Practical Knowledge Management, Knowledge based Medical Applications, Implementation Issues of Knowledge Management*

## 1 Introduction

Knowledge Management (KM) is a concept of business administration and computer science, but it is not limited to these areas, and it brings interesting ideas which could be useful for other sciences. There are various definitions and discussions on what exactly KM is about [1, 3, 4, 5, and 6], the definition of KM and Knowledge Based Systems (KBS) are interconnected to the definition of knowledge. In practice, the terms knowledge, information and data are often used interchangeably, but giving a correct definition of knowledge is possible by distinguishing among knowledge, information and data [3].
A commonly held view with sundry minor variants is that data is raw numbers and facts, information is processed data, and knowledge is authenticated information [9 and 11]. In fact Data is unclassified and unprocessed values, a static set of transactional elements, such as 211102345, John is 6 feet tall, or structured records of transactions. Information is meaningful collection of data which has been processed and organized. Knowledge is a set of understandings and the state of knowing acquireed through experience or study to use in decision making activities. It includes facts, opinions, ideas, theories, principles, models, ignorance, awareness, familiarity, understanding, facility, and so on. Knowledge is based on data and Information. Information Systems (IS) can convert data into information, but it is not able to convert information to knowledge.

## 2 KM and KBS

As we mentioned before, definition of KM and KBS depend on knowledge. Alavi and Leidner [3] have defined five different perspectives for knowledge. KM involves enhancing individual's learning and understanding through provision of information. It is the process of creation, sharing, and distributing knowledge and to build and manage knowledge stocks. KM is also organized access to and retrieval of content. t is about building core competencies and understanding strategic know-how, too. [3]

The role of IT in KBSs is to provide access to sources of knowledge rather than knowledge itself, gathering, storing, and transferring knowledge and to provide link among sources of knowledge to create wider breadth and depth of knowledge flows. Providing effective search and retrieval mechanisms for locating relevant information is also supported with IT in KBSs. IT can enhance intellectual capital by supporting development of individual and organizational competencies in KBSs [3]. Generally KM is any attempt to convert employees' knowledge (personal properties) to shared organizational knowledge (organizational property). It is aim to improve effectiveness and speedup, and decrease expenses of an organization. Giving correct knowledge in the suitable format and time to a proper expert or individual with reasonable costs is the aim of KM. Therefore:

Knowledge is explicit and/or tacit, individual and/or organizational and authenticated information which has been acquireed through study or experience to potentially influence action, and applying expertise.

KM is the practice of collecting knowledge and selectively applying it from previous experiences of decision making to current and future decision making activities with the express purpose of improving the organization's effectiveness [6].

KMS is any system which is aiming to improve the effectiveness and speedup of the organization based on the previous experiences and activities. KBS is able to create new knowledge, exchange the current knowledge, and apply all to the system.

Ann Macintosh of the Artificial Intelligence Applications Institute has written a "Position Paper on Knowledge Asset Management" that identifies some of specific business factors about necessarily of KM in contemporary life, including:

- Marketplaces are increasingly competitive and the rate of innovation is rising.

- Reduce of staffs create a need to replace informal knowledge with formal methods.
- Competitiveness reduces the size of the work force that holds valuable business knowledge.
- The amount of time available to experience and acquire knowledge has diminished.
- Early requirements and increasing mobility of work forces lead to loss of knowledge.
- There is a need to manage increasing complexity as small operating companies are transnationals sourcing operations.
- Change in strategic direction may result in the loss of knowledge in a specific area.

As a result KM enables sharing of essential knowledge to complete organization tasks and provides improvements in organizations quality and human performance and competitive advantage.

### 2.1 Practical issues and challenges of KM

KM is being used in every science, industry and business. If developers and designers of KBS don't have enough information about implementation requirements of KM and KBS, their implementation results will not be as successful. In the following section we present some success factors and common failures of KM implementations.

Any failure and mistake of individuals has a cost to the system. Therefore in case of any failure, the system should analyse and predicate enough information about the failure for the future scopes. By transferring acquired knowledge of failures to the current and incoming individuals, the amount of failures will be decreased in the system and therefore KM will succeed in decreasing the costs of organization.

Feedback data from customers is very important in optimizing future generations of products and quality management. Therefore the system should provide a completely creative backbone to collect feedback data and keep a close relationship with customers and users. Any missing feedback data leads to unsuccessful scenarios of KM in optimizing system and quality issues. After obtaining the feedback data, the system should analyse, abstract, and preserve it.

Due to staff awareness of former activities and projects some experiments are repeated in the system. As a result informing individuals about former works and projects in the organization is required. The progress of former experiments and results should be stored in a knowledge repository of the system and should be easily accessible to the individuals of the organization at any time.

Due to security and trust issues, thoughts, ideas, and good functionalities are not often shared with expert. Successful KM implementation needs to create the trustworthy culture and teamwork in which professionals feel safe and secure, and can trust in the system; providing the safe and trustable culture is critical to KM success in any system. Designers and administrators' awareness of basic psychological rules lead to the defeat of KBS. Sharing of key information and progresses needs individuals' trust.

Only a limited amount of people in the organizations are aware of key knowledge about their projects and works. That is because only a few people work in the key positions of projects, therefore when these people are absent, others miss key knowledge and it is more cost-intensive for an organization to reacquire the missed knowledge. When sharing of key knowledge is not possible generally, system should provide a secure process to preserve key knowledge. Security techniques such as encryption methods should be used to make it more secure. Access to this knowledge should be controlled securely and the system should allow the use of them just in specific cases, but not generally.

Organizational Learning (OL) is always slow. Resulting inconveniences include delays in development of products or missing better opportunities. Therefore organizations need to employ better and more rapid technologies to accelerate their learning methods. Because of great developments in IT and their application in learning methods, it could be useful to optimize OL. Members of an organization will be disappointed when they couldn't acquire knowledge from the system. Therefore providing good access to the available knowledge resources is another requirement of KBSs.

In addition to mentioned propositions, some other success factors of KM implementation are as follows:
- Employee involvement
- Information System (IS) infrastructure
- Performance measurement
- Benchmarking
- Knowledge-friendly culture
- Optimized knowledge representation methods
- Knowledge abstraction usage
- Intelligent data mining methods
- Organizational culture
- Motivational incentives for KM users
- A standard and flexible knowledge structure
- Ability to transfer knowledge with another KBSs via recent and IT technologies like Web 2.0 or Cloud computing
- Security and knowledge protection

## 3 KM in Medical Science

Various fields of application, huge amounts of data and experts with specific know-how are the main reasons why the field of medical research can be seen as a precursor for other branches of applied KM. Evolutionary algorithms, fuzzy logic or neural networks have the potential of improving knowledge and can help medical practitioners in creating diagnoses for different diseases.

Conducting knowledge audits are necessary to determine specific experts in a healthcare organization who have knowledge on a certain issue (e.g. disease or therapy). This process called "Knowledge Identification and Capture" and is one of the important steps in Medical KM (MKM).

In the healthcare setting, knowledge creation can take place in terms of improved organizational processes and systems in hospitals, advances in medical methods and therapies, better patient relationship management practices, and improved ways of working within the healthcare organization.

A healthcare organization is a collection of professional specialists who contribute to the delivery of patient care, but also often act competitively inside the organization, without being willing to transfer knowledge because of associated status and power within the organization and the society. Effective knowledge management in the Medical Application area as well as in other branches requires a "knowledge sharing" culture to be successful.

Knowledge application refers to taking the shared knowledge and internalizing it within one's perspective and worldviews. Knowledge-enabling technologies which can effectively be applied to healthcare organizations are: Groupware, Intranet, Collaborative tools and Knowledge Portals.

### 3.1 Knowledge based Medical Systems

By incorporating applied KM and medical science many fields of application can be discovered. Modern medicine generates huge amounts of heterogeneous data [14] almost daily. Researchers and medical practitioners have to deal with data on one hand and with personnel know-how on the other hand. Knowledge discovery as an early step in the KM process transforms data into knowledge [14], and helps medical practitioners in that way to decrease information overload. Knowledge Management and Medical Engineering center (KMME) in corporation with some medical experts and clinical centers discovered a wide gap between raw data and concluded knowledge. Disparity of data collection and data comprehension makes computerized techniques necessary to help humans to address this problem [14]. In the following section three projects of the KMME and one another MKM project will be described briefly with their aims and benefits.

#### 3.1.1 StroPoS

The Stroke Portal System for emergency risk patients (StroPoS) was implemented based on the concept of a corresponding portal solution [21]. The need for a portal like StroPoS results from serious influences strokes have on the association. Strokes are the fifth common cause of death in Germany in 2008 [25] and they are often affiliated with severe disablements. Early diagnosis supported by a Stroke Portal System could decrease the costs of therapy and aftercare immensely.

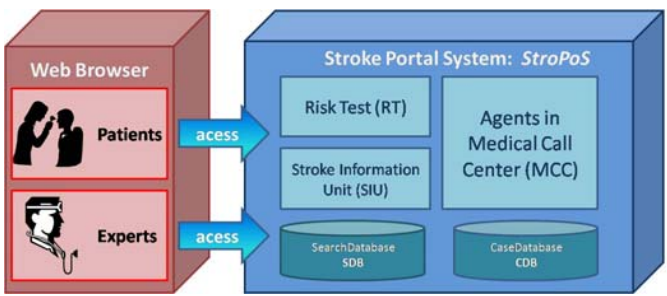

Figure 1: Architecture of the Stroke Portal StroPoS

The architecture of StroPoS System has been illustrated in Figure 1. Furthermore, a Stroke Portal System should offer a service which recognizes warnings. Concerning recognizing warnings, a Medical Call Center (MCC) with agents can help as well. The KMME developed several methods of KM within this portal which will be integrated and realized in the next steps of the project.

Patients and experts can access StroPoS via Web browser. Some other components like the Risk Test (RT) and Stroke Information Unit (SIU) can be accessed by the MCC. The system is divided into two parts: Patient section and expert section. While patients have free access to information like definition and descriptions about the stroke disease, experts have to log in to get access to the expert area on the portal. After this authentication procedure they can use the CDB, which includes patient case studies, as well as the SDB for DB queries. In the database experts can find special stroke cases and can set up several parameters to localize a case. Experts can also make an anonymous case file and upload it to share with other experts. The challenge was to implement a program structure by abstracting the anonymous example health records for entering, saving and comparing different courses of disease [20]. With the RT, users can test their stroke risk. The test contains ten questions which are easy to understand and be proved by clinical partners of KBS group.

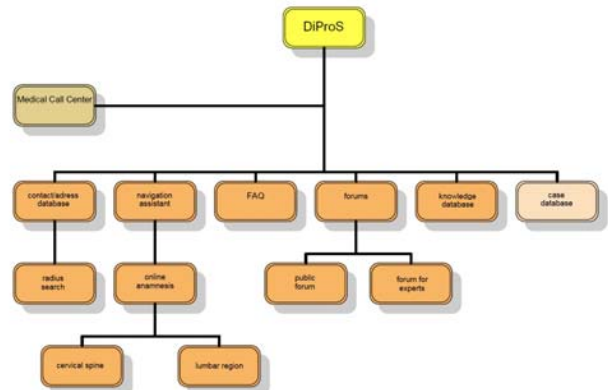

Figure 2: Diagram of the DiProS project

Users have also the opportunity to search for special stroke clinics. To find the nearest clinic users need only to type their post code, and system will find nearest stroke clinic.

#### 3.1.2 DiProS

Disc Prolapse System (DiProS) is another knowledge-based medical application implemented by KMME at the University of Siegen. The major goal of the project is to achieve a modular and structural portal in the field of disc herniation disease. DiProS focuses on allocation of reliable and authentic information supplemented by additional knowledge for experts and individuals, regardless if they are just interested or affected by spinal disc herniation. The offered information in the portal is extended by interactive services, which allow instantaneous contact to medical practitioners and make telemedical consultation possible. Methods for estimation of individual patients are also offered in the portal.

The reuse of components like the CBD or the forums saved a lot of time but it also shows that the components can be reused in other projects.

Besides the well-proven components, new components have been developed, for instance, the navigation assistant which helps users to navigate through the system and find the information they need.

### 3.1.3 AlWiP

This project is another implemented project by KMME. The major goal of the Alzheimer Wissen Portal (AlWiP) project is the early detection of Alzheimer disease indicators. Reasons, symptoms and factors of risk have to be evaluated through data and picture material (i.e. brain MRT's, course of brain alteration) in terms of correlation and integrated into the Knowledge portal as a decision support component.

The whole project aims to create a knowledge management portal for the medical application of Alzheimer's disease. An image analysis of the brain, combined with information about the general state of health of patients can indicate the danger of developing Alzheimer's. (Fig. 3)

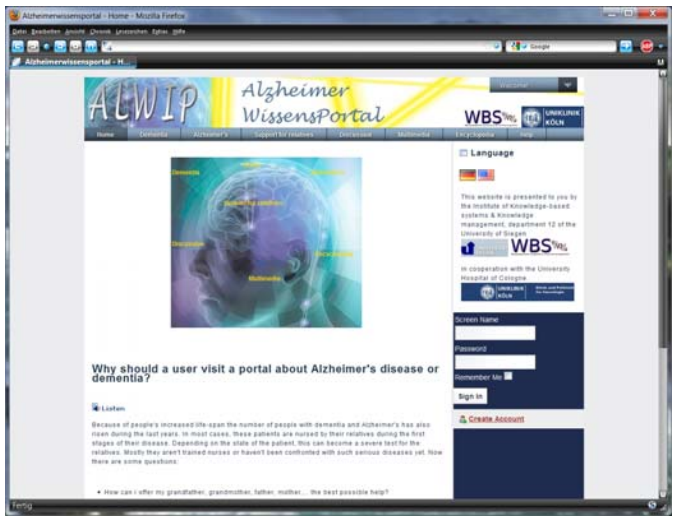

Figure 3: Alzheimer Knowledge Portal

With the help of the Alzheimer's and dementia specialists at the University of Cologne (the cooperation partners of the KMME), an Alzheimer's knowledge database was built for users such as family members of patients and experts, in which different forms of dementia were shown.

In the discussion area, family members are able to share their experience with other concerned persons. In this way people can share their knowledge and find solution for their special problems.

The portal should not replace other information sources like health personnel, expert's guides [18] or caregiver's guides [7]. In fact the portal is planned as a knowledge platform which completes these traditional resources. Beside this AlWiP is a quick and easy opportunity to get Alzheimer information on demand.

For best possible user support a navigation assistant is implemented in the portal. It enables the user to ask questions like 'what is dementia'. Another new feature is that if there is reason why a user cannot read articles, the assistant can read out articles and every text file on the portal Therefore it is possible for that person to receive needed information by using the assistant.

### 3.1.4 OncoDoc

Developed and implemented in the Service d'Oncologie Medicale Pitie-Salpetriere (Paris, France) [12], OncoDoc is a decision support system designed to provide best therapeutic recommendations for breast cancer patients. While clinical trials offer cancer patients the optimum treatment, historical accrual of such patients has not been very successful. Developed as a browsing tool of a knowledge base structured as a decision tree, OncoDoc allows physicians to control the contextual instantiation of patient characteristics to build the best formal equivalent of an actual patient [26].

Originally published as textual documents, clinical practice guidelines have poorly penetrated medical practice because their editorial properties do not allow the reader to easily solve a given medical problem at the point of care [15]. OncoDoc has been developed as a guideline-based decision support system to provide best treatment recommendations [15]. It is non-automated and allows flexibility in guideline interpretation to obtain best patient-specific recommendations at the point of care [24]. Rather than providing automated decision support, OncoDoc allows medical practitioners to control the operationalization of guideline knowledge through his hypertextual reading of a knowledge base encoded as a decision tree. In this way, physicians have the opportunity to interpret the information provided in the context of the patient, therefore, controlling the categorization to the closest matching formal patient [15].

Within the tree different parameters describe the actual state of a patient. The complete expansion of the decision tree can be seen as a patient centered repository of all theoretical clinical situations that could be met within breast cancer pathology [26]. Recommendations based on the clinical state of the patient are displayed at the lower side of the decision tree and the path through it represents the clinical profile of the patient. (Fig. 4)

The System has been tested at the Institut Gustave Roussy (IGR) with a before-after study in which treatment decisions for breast cancer patients were measured before and after using the system in order to evaluate its impact upon physicians' prescribing behaviour [24]. To assess how the system could be reused in another institution which was not involved in the development process, a new experiment at IGR was conducted. Minor site-specific customizations of the knowledge base were performed. After four months, 127 cases were recorded. Results showed that there was no significant difference of physician compliance with OncoDoc (85%) when site-specific recommendations were, or not, available, although local recommendations were chosen preferably (55%), thus legitimating the adaptation [12].

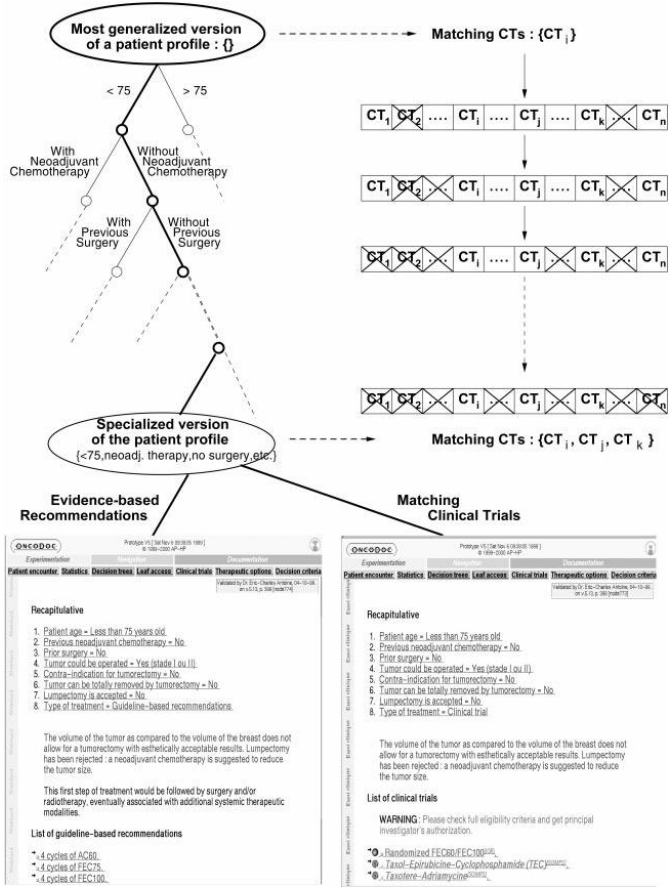

Figure 4: Representation of the process underlying eligibility criteria screening [26]

Summing up, the system uses patient characteristics to discriminate between different treatments alternatives, furthermore it offers matching clinical trials for the actual case. Physicians can be supported in disseminating between next therapy steps.

## 4 Conclusions

The domain and benefit of KM is not limited to a specific science or industry, its techniques are being implemented in each field and industry. For example, every company wants to acquire knowledge with less time and cost, and most research and innovation is based on traditional knowledge and information. All of these are possible with only the use of KM systems and techniques.

After studying different projects and their focus we conclude that KM is going to be more important and useful in medical science and assisted living, implementing knowledge based portals and knowledge learning techniques for future innovations. Of course KM is not limited to medical science, but it has had very sensitive growth and success in this area. Most medical centers and doctors want to function in real time and medicate their patients immediately, therefore research in assisted living projects is the most highlighted and this could be more efficient by employment of KM.

First minutes of most medical attacks are very important for the consequences of disease. When patients are living in their private areas, direct support of the patient during the first minutes of an attack is not possible for the medical and support team. Therefore having urgent remote medicating and guidance system could be more helpful to decrease the consequences of diseases. That is the basic idea of assisted living projects which is common in most research institutes nowadays. Since some diseases need full time controlling and it is not possible for everybody to employ a full time nurse, the importance of remote and intelligent controlling systems are clear for everybody. One example of these applications could be knowledge based portals. They are able to share the knowledge of experts with individuals, exchange professional knowledge between experts, find the best fit and nearest medical centers for patients, find and introduce special cases of diseases for experts, establish panel discussion between experts and individuals, and so many other facilities. All of these services use different categories and techniques of KM. As a result KM is used in most sciences and industries because of its reasonable advantages to the systems, but nowadays it is more successful and helpful in medical sciences and most medical centers are interested in implementation and using KBSs. In this paper we have studied some failures in implementation of KM in organizations and some case studies of KM in medical science. Of course we know that challenges and success factors of KM implementations are not just limited to the what we have mentioned, but these are the most basic and mandatory conditions which every KBS should support. We think that KM will be more innovative and important given time.